# SOLUTION OF THE ONE-DIMENSIONAL BOND PROBLEM IN A PERCOLATION THEORY

M.A. Bureeva, V.N. Udodov

The results of investigations of main characteristics of a one-dimensional percolation theory (percolation threshold, critical exponents of correlation radius and specific heat, and free energy) are presented for the problem of bonds and sites. For the first time it is shown that for a finite-size system the stability condition is fulfilled while the scaling hypothesis is inacceptable for one-dimensional bond problem.

## INTRODUCTION

The percolation theory is an intensively developing field of the physics of randomly inhomogeneous media [1–4]. It involves formulation of the following problems:
- Connection of a very large (macroscopic) number of elements (atoms, molecules, grains, atomic layers, etc.), given that a bond of every element with its neighbors is a stochastic quantity which is derived in a certain way;
- Laws controlling the behavior of different quantities in the vicinity of the percolation threshold, which is similar to the point of an ordinary phase transition (PT).

A comparatively new field of application of the percolation theory is simulation of polytypic transformations [1–4]. It is well known that a polytypic close-packed crystal can be presented using the Ising model. To describe a real 3-D close-packed crystal within an axial Ising model, use is generally made of a one-dimensional lattice gas model [4].

According to the lattice gas model, the sites are filled with zeros and unities. Note that any sequence of zeros and unities can compose a set of 3C (FCC), 4H (double HCP), and 2H (HCP) structures separated by stacking faults. The lengths of sequence sections correspond to the thicknesses of crystal blocks. The lattice gas is then investigated by the Monte Carlo method within the percolation approach [2]. Two more fields of application of the one-dimensional percolation theory should be mentioned: anomalous diffusion in one-dimensional (quasi-one-dimensional) defect structures [3] and quasi-one-dimensional Ising magnetics with non-magnetic atoms [5]. The problem of electrical conduction of doped semiconductors can be reduced to the study of diffusion of charged particles within the percolation approach [6–9].

Despite the interest in the one-dimensional percolation theory, the available literature contains little on the values of its characteristics (percolation threshold, critical exponents) as a function of the chain length; this particularly concerns the bond problem. In this work we performed an investigation of the main characteristics of the one-dimensional percolation theory (percolation threshold, critical exponents of correlation radius and specific heat and free energy) for the bond and site problems for the percolation radii $R = 2$ and 3 for finite-size crystals (up to hundreds of sites).

## 1. MODEL

A model chain of sites was used to solve the bond and site problems [6–13]. The site problem is reduced to the following. All bonds in the chain are considered to be whole, and the sites overlap in a stochastic fashion. Two whole sites are connected if they are located next to each other or separated by a chain whole sites, between which only single blocked 0sites occur

(percolation radius is equal to 2) [10, 12, 13]. The radius equal to 3 is determined in a similar manner.

A cluster in this context will be an aggregate of connected whole sites. A subtending (connecting, infinite) cluster will be the one connecting the opposite sides of the lattice [6, 7]. Percolation is said to occur in the case where a connecting cluster is present. A correlation length $\xi$ is the mean maximum length of a non-connecting cluster [6, 7]. For a certain critical value of the fraction of whole sites $x_c$, the probability of existence of a connecting cluster vanishes. In other words, as the fraction of whole sites $x$ decreases, percolation from one end of the chain to the other at this value of $x_c$ vanishes. This value of $x_c$ is referred to as the percolation threshold. It turns out that there is a subtending cluster only for $x > x_c$. For $x < x_c$, there are only small clusters. The threshold $x_c$ corresponds to the point of an ordinary PT [1, 4, 6–16].

If we use $N$ to denote the number of sites in the sequence and $N_1$ – the number of whole sites in the chain at the percolation threshold, then the percolation threshold will be given by

$$x_c = \frac{N_1}{N}. \tag{1}$$

In the bond problem, on the contrary, all the sites are whole, and the bonds are broken randomly and can be whole or blocked [6, 7]. The bond problem is reduced to that of sites, but is solved on a different lattice termed covering. A covering lattice is constructed from the initial (containing) lattice, using the following rules [6, 7]: 1) in the middle of every bond of the initial lattice a site of the covering lattice is built, 2) two sites of the covering lattice are connected with each other in the case and only in the case where the bonds of the initial lattice, whereon these two sites are built, converge in the site of the initial lattice. This yields us a new periodic lattice that is referred to as a covering lattice with respect to the containing (initial) lattice. Hence it follows that the percolation threshold of the bond problem on the initial lattice is equal to that in the site problem on the covering lattice. Hammersley proved [6] that for any lattice (not necessarily plane) the threshold value for the bond problem is not higher than that in the sites problem

$$x_{cb} \leq x_{cs}, \tag{2}$$

where $x_{cb}$ is the percolation threshold in the bond problem and $x_{cs}$ is the percolation threshold in the site problem. Inequality (2) will be referred to here as the Hammersley theorem [6].
In order to find the percolation threshold and the critical exponents of the bond problem, use is made of the methods of computer simulation [6–16]. Construct a one-dimensional array whose elements would be numbers 0 (blocked bond) and 1 (whole bond). The initial parameters would be the length of the initial chain, the percolation threshold and the number of runs. The length of the initial chain is $5 \leq N \leq 300$. Since we now consider the bond problem, the number of elements in the array would be equal to the number of bonds in the problem, which is found using the following formula [10]:

$$n = \frac{R(2N + R - 3)}{2}, \tag{3}$$

where $R$ is the percolation radius.

The percolation radius values are $R = 2$ and 3. The number of runs is $m = 10\,000$. First, we consider all the bonds to be whole (array element values are equal to 1).

In addition to finding the percolation threshold, the following characteristics are to be calculated [7, 9, 11]:

$s$ – cluster size, the number of whole sites in a cluster,

$\langle N_S \rangle$ – average number of clusters of size $S$,

$n_S(x) = \dfrac{\langle N_S \rangle}{N}$ – cluster size distribution,

$P_\infty(x) = \dfrac{M(N)}{N}$ – percolation probability – probability that a randomly chosen site belongs to the subtending cluster ($M(N)$ – average size of the largest cluster; the largest cluster is the one containing a maximum number of whole sites).

Since the percolation theory describes a geometrical phase transition (PT), some of its quantities are similar to those of the PT theory of the second order. For instance, within the percolation theory the concentration of whole sites plays the same role as does the temperature in a temperature-induced PT. The probability that a site belongs to an infinite cluster (percolation probability) is similar to the order parameter in the theory of temperature-induced PTs. Many important cluster characteristics (correlation length, average number of sites) in the vicinity of the transition are described by the exponential function with different critical exponents [11]

$$F(x) \propto |x - x_c|^{2-\alpha},$$

$$P_\infty(x) = \sum_s s\, n_S(x) \propto |x - x_c|^\beta,$$

$$\xi(x) \propto (x_c - x)^{-\nu}.$$

In this work, using the methods of computer simulation, we have found the values of percolation threshold ($x_c$), free energy, critical exponents of the correlation radius $\nu$ and specific heat $\alpha$ in one-dimensional site and bond problems for chains of different lengths for the percolation radius values $R = 2$ and 3. In addition, we have verified the validity of the Hammersley theorem and the scaling hypothesis for a finite-size system.

## 2. RESULTS OF INVESTIGATION

We have calculated the percolation threshold in a one-dimensional problem of sites and the problem of bonds for different chains (from 3 to 300 sites), given that the percolation radius $R = 2$ (Fig. 1).

It is evident from Fig. 1 (curves *1* and *2a*), that the calculated percolation threshold values satisfy the Hammersley theorem [6–10, 17].

Furthermore, we have calculated the percolation threshold in a one-dimensional bond problem for different chains (from 3 to 300 sites) for the percolation radius values $R = 2$ and 3 (Fig. 1, curves *2a* and *2b*). It was revealed that with increase in the percolation radius the percolation threshold decreases, but, nonetheless, tends to unity for $N \to \infty$, where $N$ is the length of the chains.

The relative calculation error was found to be within 7–25%, decreasing with increase in the length of the chains [17].

The critical exponent of the correlation length $\nu$ was calculated from the relation [3, 5]

$$\delta(N) \propto N^{-\frac{1}{d\nu}}, \tag{5}$$

where $\delta(N)$ is the rms deviation of the percolation threshold from the mean threshold value, $d$ is the spatial dimension (in our case $d = 1$).

Exponent $\nu$ decreases as the length of the sequence is increased but increases with the percolation radius (for sufficiently long chains) (Fig. 2).

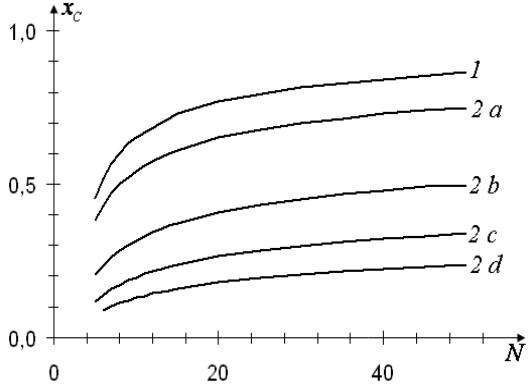
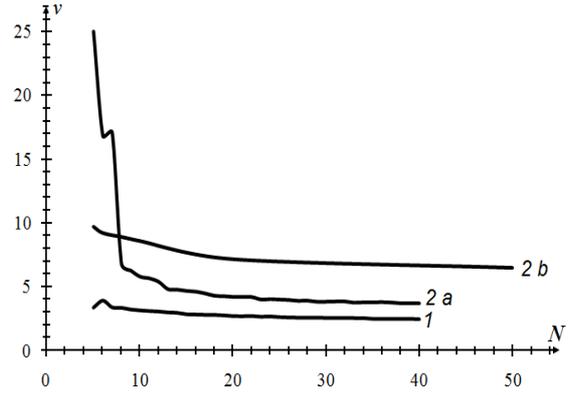

Fig. 1. Percolation threshold dependence in one-dimensional problems of bonds and sites on the chain length: site problem (curve *1*, $R = 2$) bond problem (curve *2a*, $R = 2$), bond problem (curve *2b*, $R = 3$), bond problem (curve *2c*, $R = 4$)б bond problem (curve *2в*, $R = 5$). Relative error is 7–25 %.

Fig. 2. Dependence of exponent ν in one-dimensional problems of bonds and sites on the chain length: site problem (curve *1*, $R = 2$), bond problem (curve *2a*, $R = 2$), bond problem (curve *2b*, $R = 3$). Relative error is 2–5% in the site problem and 20–25% in the bond problem.

Note that the correlation length exponent in the one-dimensional percolation theory is anomalously large compared to the 2-D and 3-D cases for ordinary phase transitions. Recall that in the Landau PT theory [14] the correlation length exponent is 0.5. It is especially large in the bond problem (Fig. 2) [17]. This implies that the correlation length dependence on temperature in the one-dimensional percolation theory for nanosized systems is by far stronger than it is in the ordinary phase transitions in macroscopic systems.

It is the average number of clusters calculated per single site, which plays the role of free energy $F$ in the problem under study [8, 9]:

$$F(x) = \sum_s n_s e^{-sh}, \qquad (6)$$

where $h$ is the external field.

In the site problem $x$ is an analog of temperature (or pressure) for an ordinary PT. From the physics standpoint, this free energy in the problem of close-packed polytypes is associated with the shuffling of close-packed layers with respect o each other [1, 4].

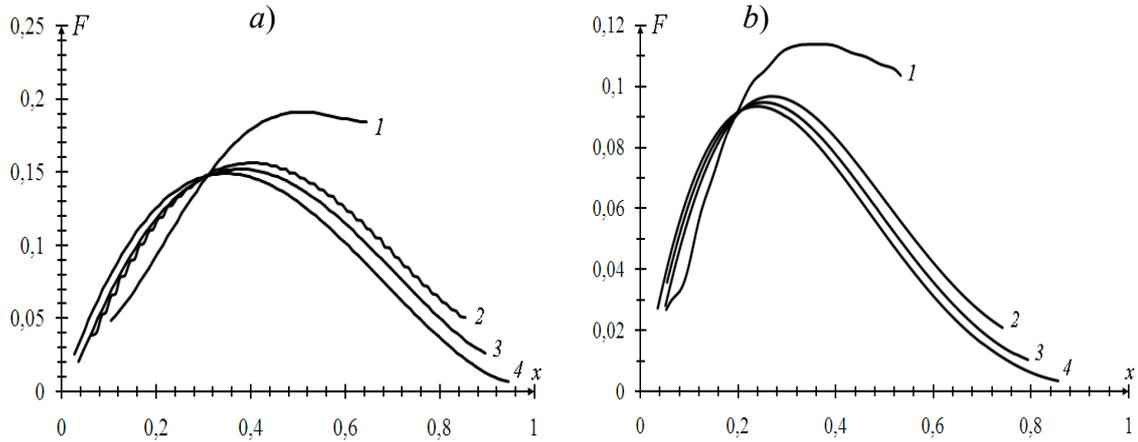

Fig. 3. Free energy dependence on the chain length and whole-site fraction in the problems of sites (*a*) and bonds (*b*), $h = 0$, $R = 2$. Number of sites in a sequence: $N = 10$ (curve *1*), 50 (curve *2*), 100 (curve *3*) and 400 (curve *4*) (*a*), $N = 10$ (curve *1*), 50 (curve *2*), 100 (curve *3*) and 300 (curve *4*) (*b*). Relative error: 0.3–1% (*a*) and 1–5% (*b*).

We have also calculated the values of free energy in a one-dimensional problem of sites and bonds for a chains of different length for an external field of $h = 0$ (Fig. 3). The resulting dependence in both cases is a convex curve, which agrees with the thermodynamic stability conditions of the system [8, 9, 17]. It should be also noted that the free energy maximum in the bond problem is shifted to the left with respect to the site problem, implying that it is achieved for a smaller number of whole sites in the chains.

With increase in the external field, free energy also decreases (Fig. 4) [11].

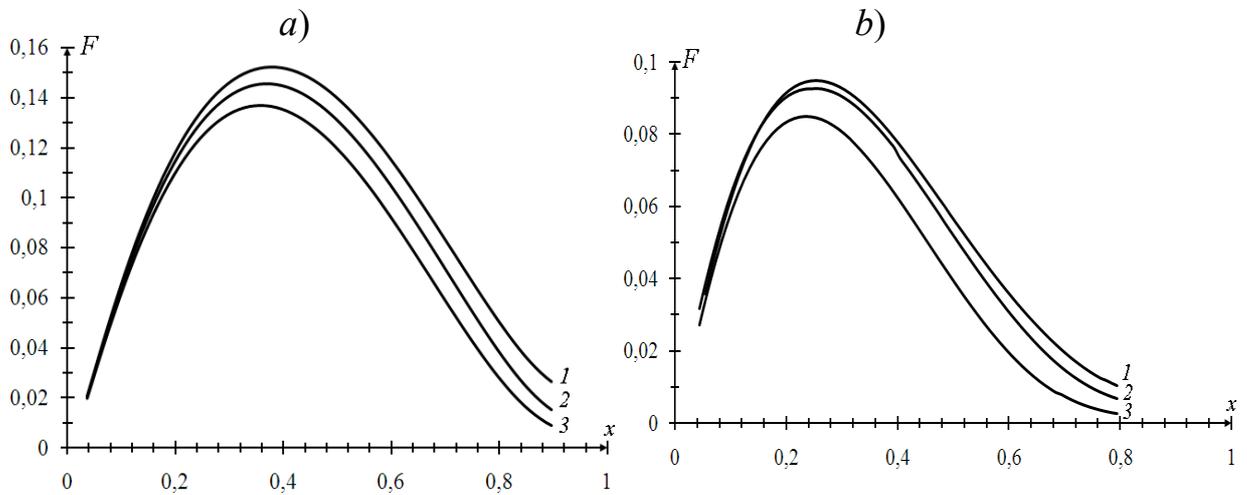

Fig. 4. Dependence of free energy per site on the external field and the fraction of whole sites in the site problem (a) and bond problem (b): $R = 2$, chain length $N = 100$. External field: $h = 0$ (curve *1*), 0.02 (curve *2*) and 0.05 (curve *3*) (*a*), $h = 0$ (curve *1*), 0.01 (curve *2*) and 0.05 (curve *3*) (*b*). Relative error: 0.5–0.7% (*a*) and 1–8% (*b*).

The critical exponent of specific heat α in a weak external field (Fig. 5) was found from the relation [8, 9]

$$F \propto |x - x_c|^{2-\alpha}, \qquad (7)$$

where $F$ is the free energy, $x < x_c$, $x \to x_c$ (Fig. 5). Both in the site problem and in the bond problem, exponent α decreases with increasing the chain length and the external field value [10].

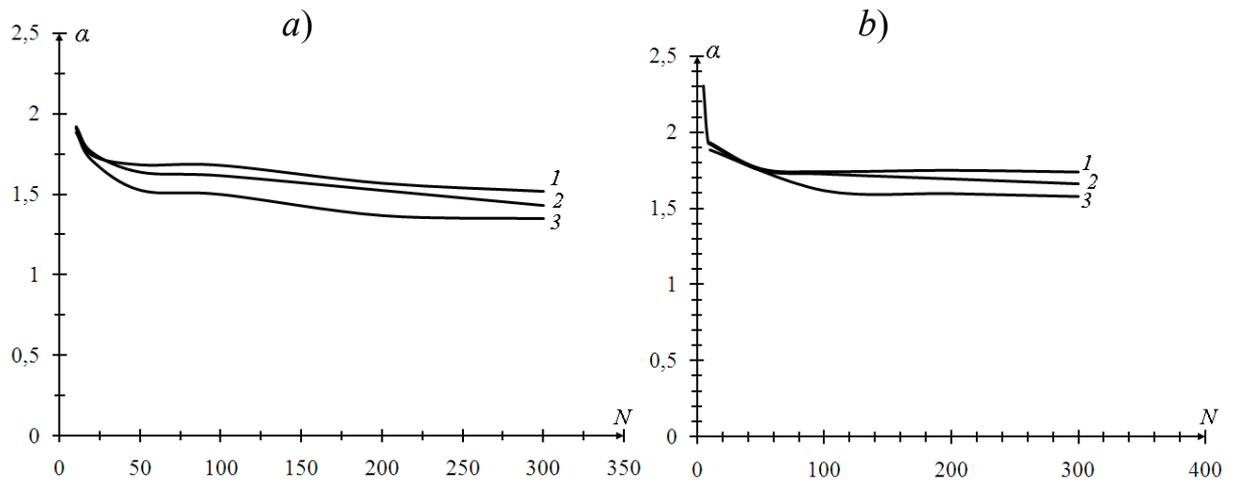

Fig. 5. Dependence of critical exponent α on the chain length and the external field value in the problems of sites (*a*) and bonds (*b*), $R = 2$. External field: $h = 0$ (curve *1*), 0.02 (curve *2*) and 0.05 (curve *3*) (*a*), $h = 0$ (curve *1*), 0.01 (curve *2*) and 0.05 (curve *3*). Relative error: 2–4% (*a*) and 2–8% (*b*).

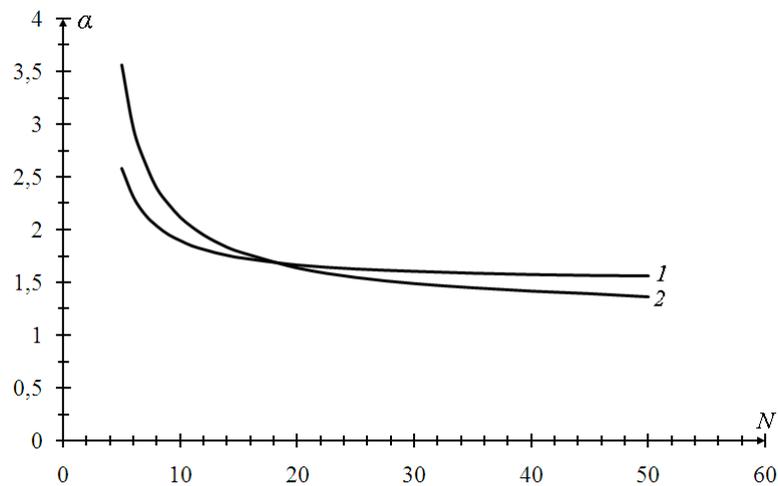

Fig. 6. Dependence of critical exponent of specific heat in a one-dimensional problem of bonds on the chain length, $R = 2$ (curve *1*) and 3 (curve *2*). Relative error: 2–10%.

It has been also revealed that the critical exponent of specific heat in the bond problem is found to increase with the percolation radius (Fig. 6) [13, 17].

It has been shown that the stability condition $\alpha + vd \geq 2$ ($d$ is the spatial dimension, in our case $d = 1$, and $v$ is the critical exponent of the correlation length [13]) is always fulfilled as an inequality.

Table1. Critical exponents in one-dimensional problems of sites and bonds (percolation radius $R = 2$)

| Site problem | | | | Bond problem | | | |
|---|---|---|---|---|---|---|---|
| $N$ | $\alpha$ | $\nu$ | $\alpha + \nu d$ | $N$ | $\alpha$ | $\nu$ | $\alpha + \nu d$ |
| 5 | 2.48081 | 3.3187 | 5.79951 | 5 | 2.30345 | 24.9637 | 27.26715 |
| 10 | 1.93111 | 3.0302 | 4.96131 | 10 | 1.93418 | 1.9411 | 3.87528 |
| 20 | 1.80695 | 2.7445 | 4.55145 | 20 | 1.80674 | 4.1951 | 6.00184 |
| 50 | 1.734 | 2.7002 | 4.4342 | 50 | 1.76429 | 3.4491 | 5.21339 |
| 100 | 1.7591 | 2.6458 | 4.4049 | 100 | 1.74426 | 3.2843 | 5.02856 |

It is evident from Table 1 that as the chain length is increased, the value of $\alpha + \nu d$ decreases, but is always larger than two. An extrapolation with respect to the inverse size shows that for an infinite system in a one-dimensional problem of bonds $\alpha + \nu d \approx 4{,}8$. Thus, the scaling hypothesis [6–9] for a finite-size system (and, according to the estimates, even for an infinite-size system) within the one-dimensional percolation theory is violated both in the problem of sites and bonds. Recall that according to the scaling hypothesis, the inequality $\alpha + \nu d \geq 2$ (stability condition) is degenerated into equality [8, 9]. This violation turns out to be so significant that we have to draw a conclusion on inapplicability of the scaling hypothesis to random one-dimensional nanostructures measuring hundreds and thousands structural elements. The corollaries of this conclusion require further investigation.

**SUMMARY**

1. A new algorithm has been proposed, which simplifies calculation of the percolation threshold in a one-dimensional bond problem for finite-length chain by the methods of computer simulation.

2. The percolation threshold has been calculated in a one-dimensional bond problem for chains of various lengths (from 3 to 300 sites) for the percolation radius values $R = 2$ and 3. It has been shown that the percolation threshold values in the bond and site problems in one-dimensional finite-size crystals satisfy the Hammersley theorem.

3. The values of the correlation length exponent have been calculated in one-dimensional problems of bonds and sites for the percolation radius values $R = 2$ and 3 for the chains containing up to 300 sites. The correlation length exponent $\nu$ in a one-dimensional problem of bonds has been found to exceed the values of $\nu$ in the problem of sites for equal-length chains for the percolation radius $R = 2$, and, in general, this exponent was found to be extraordinary large compared to the 2-D and 3-D cases for ordinary phase transitions in macrosystems.

4. An analog of free energy and a critical exponent of the specific heat analog in the site and bond problems have been calculated within the percolation theory for varying length chains (from 10 to 400 sites) for different values of the external field.

5. It has been found out that the values of the critical exponent of the specific heat analog in one-dimensional problems of bonds and sites decreases as the length of crystal is increased, given that the thermodynamic stability condition is fulfilled.

6. It has been demonstrated that with increase in the percolation threshold in the bond problem, the percolation threshold and free energy decrease while the critical exponents of the correlation length $\nu$ and those of the specific heat analog $\alpha$ are increased. In the problem of close-packed polytypes, the above critical exponents are associated with the shuffling of close-packed planes of a real three-dimensional crystal.

7. It has been shown that the scaling hypothesis in the model of one-dimensional percolation for the percolation radius values $R = 2$ and 3 in the problems of sites and bonds is inapplicable to a finite-size (hundreds of sites) system. Thus, the scaling hypothesis is inapplicable to random (disordered) one-dimensional nanostructures containing hundreds of structural elements. This might (to a certain extent) account for the anomalous properties of nanometer systems.

8. The results obtained in this work can be used in modeling hopping conduction in semiconductors at low temperatures [3], polytype transformations in close-packed crystals [2, 4] and in a number of other cases [7–16], especially in the case of the objects or grains of nanometer size [17].

Thus, for the first time, using the method of computer simulation, we have solved the bond problem for the model of one-dimensional percolation in finite-size systems of tens of nanometers.